\documentclass{eas}
\usepackage{graphicx}
\begin{document}
\title{
Where Does the Disk Turn Into the Halo?\\  Cool H{\tt I} in the Outer Milky Way Disk
}
\author{John M. Dickey}\address{University of Tasmania, Maths and Physics - Bag
37, Hobart, TAS 7005, Australia}
\begin{abstract}
Using H{\tt I} absorption spectra 
taken from the recent surveys of $\lambda$21-cm line and continuum emission
in the Galactic plane, the distribution of cool atomic clouds in the
outer disk of the Milky Way is revealed.  The warp of the midplane
is clearly seen in absorption, as it is in emission, and the cool,
neutral medium also shows flaring or increase in scale height with radius 
similar to that of the warm atomic hydrogen.  The mixture of phases,
as measured by the fraction of H{\tt I} in the cool clouds relative to
the total atomic hydrogen, stays nearly constant from the solar circle
out to about 25 kpc radius.  Assuming cool phase temperature $\sim 50$ K
this indicates a mixing ratio of 15\% to 20\% cool H{\tt I}, with the rest warm.
\end{abstract}
\maketitle
\runningtitle{Cool H{\tt I} in the Outer Disk}

\section{Background}

The structure of the Milky Way interstellar medium far outside the solar circle is
in some ways similar to the transition from disk to halo. 
As the stellar surface density of the disk
decreases, its gravitational potential as a function of $z$
gets smoother and shallower, so the median pressure at midplane 
must decrease.  For heating-cooling balance, when the pressure gets
low enough, less than about 300 in units of K cm$^{-3}$, then 
the cool phase of the H{\tt I} should disappear, as there is no density for
which equilibrium can be achieved for pressure below this two-phase threshold
(Wolfire {\it et al.} \cite{Wolfire_etal_1995},
\cite{Wolfire_etal_2003}).  This might lead to a decrease in 
the amount of H{\tt I} in the cool phase as $R_G$ increases, or even possibly
an abrupt edge to the cloud population at some critical $R_G$.  
Although expected, neither of these effects is seen in this data,
at least for $R_G < 25$ kpc.

\begin{figure}
\hspace*{1.3cm}\includegraphics[width=6.2cm,angle=-90]{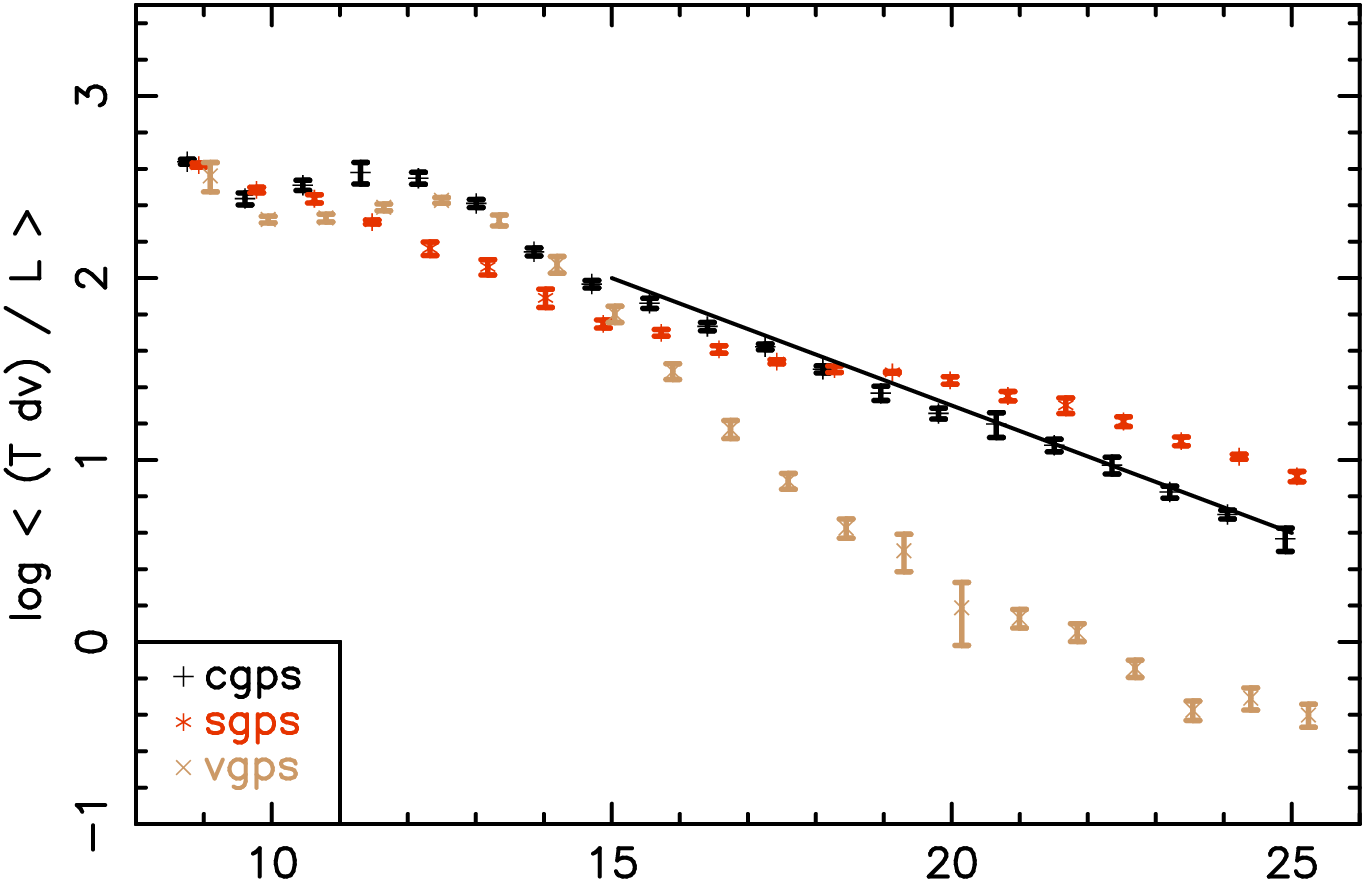}\\
\hspace*{1.2cm} \includegraphics[width=7cm,angle=-90]{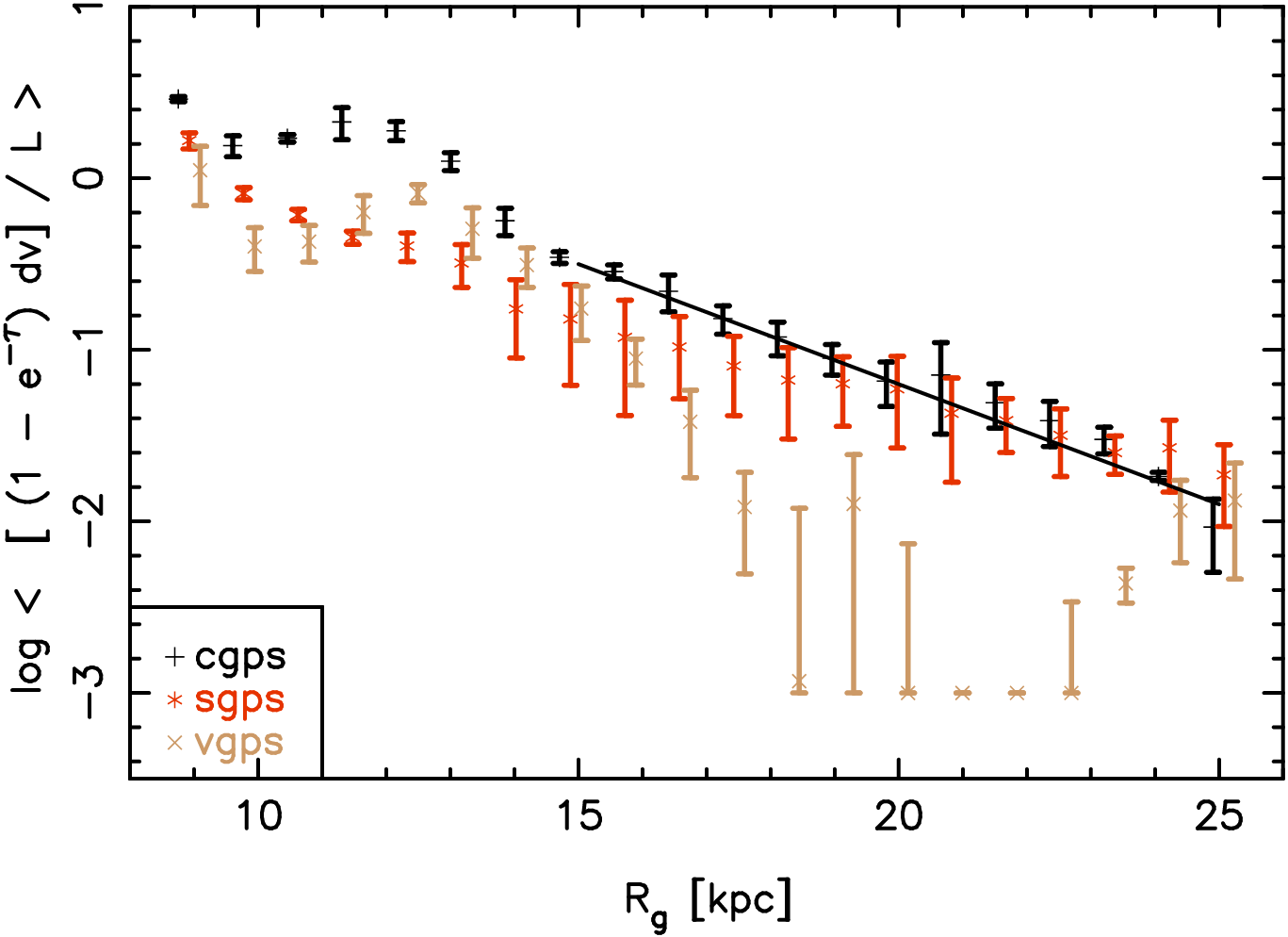}
\caption{The radial distributions of the 21-cm line emission brightness
temperature, $T$ (top), and optical depth (bottom) both plotted as 
averages per unit distance along the line of sight, {\it vs.} 
Galactocentric radius, $R_G$.  The effect of the warp in the first
quadrant is clear in the drop in both quantities in the VGPS data
at $R_G >$ 15 kpc.  The line indicates an exponential with scale
length 3.1 kpc.  For recent observational reviews of the warp and flare in the 
HI disk, see Kalberla {\it et al.} (\cite{Kalberla_etal_2007}) and
Kalberla and Kerp (\cite{Kalberla_Kerp_2009}).}
\end{figure}

\section{Data}

The absorption and emission spectra are taken from 
the Canadian Galactic Plane Survey (CGPS, Taylor {\it et al.}
\cite{Taylor_etal_2003}), the Southern Galactic
Plane Survey (SGPS, McClure-Griffiths {\it et al.}
\cite{McClure-Griffiths_etal_2005}), and the VLA Galactic Plane Survey
(VGPS, Stil {\it et al.} \cite{Stil_etal_2006}).
The methods of extracting the absorption spectra, and the
corresponding emission spectra in the directions of compact
extragalactic continuum sources, are described by Strasser
{\it et al.} (\cite{Strasser_etal_2007}).
There are some 650 such sources bright enough to give usable
absorption spectra in the three surveys.  The analysis
below uses data only in these directions,
even though there is much more emission data available, to avoid bias
that could come from using different samples for the emission
and absorption averages.  The averages are taken over three
independent samples for each survey, the first is the relatively
small number of background sources bright enough to give spectra
with optical depth noise $\sigma_{\tau} < 0.02$, the second with
$0.02 \leq \sigma_{\tau} < 0.05$, and the third with 
$0.05 \leq \sigma_{\tau} < 0.1$.  The numbers of sources in the
second and third categories are much larger, although their spectra
are relatively less reliable.  Error bars on the figures represent
the scatter among these three samples.

\begin{figure}
\hspace{.8cm}\includegraphics[width=11cm]{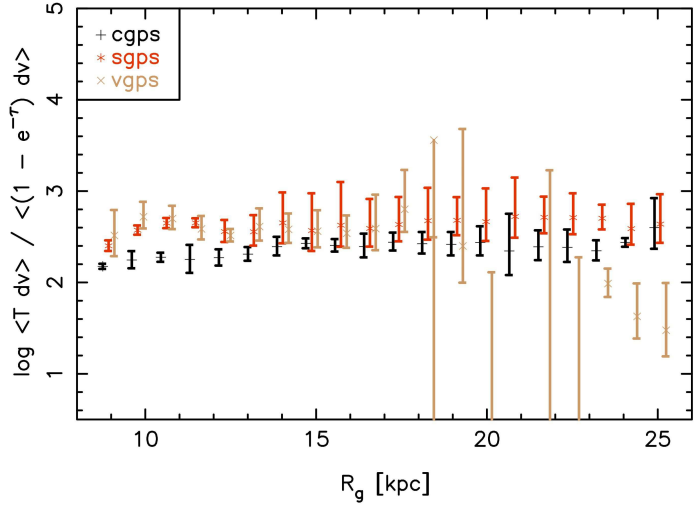}
\caption{The radial distribution of the 21-cm line excitation
temperature, $\left< T_{sp} \right>$, averaged over the different
phases.  The effect of the warp in the VGPS data (gold) at $R_G >$ 15 kpc
makes these points uncertain, as shown by the large 
error bars.}
\end{figure}

\section{Analysis}

Figure 1 shows the radial distribution of the emission (top) and
absorption (bottom), using kinematic distances based on a flat rotation curve. 
The points on figure 1 are computed by dividing the 
brightness temperature and optical depth in each spectral channel by
the line of sight path length corresponding to the velocity
width of the channel.
The brightness temperature of the 21-cm line
traces the column density of the atomic gas in the velocity
range of each spectral channel, so dividing by this
path length makes the y-axis of the
top panel proportional to the space density, $n_H$.  
Converting to cgs units makes the value 3
[meaning log (10$^3$ K km s$^{-1}$ kpc$^{-1})$]
correspond to density of 0.59 cm$^{-3}$.  The optical depth
is proportional to the average of $\frac{n_H}{T_s}$, where $T_s$ is the spin temperature,
i.e. the excitation temperature of the $\lambda$21-cm transition, which is
normally close to the kinetic temperature in the gas.
Thus the quantity displayed on the y-axis of the bottom panel of figure 1 is
proportional to the density divided by the spin
temperature, $\frac{n_H}{\left< T_{sp} \right>}$.  Here the 
average $\left< T_{sp} \right>$ means that if gas at different temperatures
contributes to the same velocity channel of the emission and absorption spectra,
then the relevant temperature for the data on figure 1 is the density
weighted harmonic mean.  Translating the y-axis of the right hand panel of figure 1
to cgs units, the value 1 [meaning log (10$^1$ km s$^{-1}$ kpc$^{-1})$] corresponds to
$\frac{n_H}{\left < T_{sp} \right>}$ = 5.9 10$^{-3}$ cm$^{-3}$ K$^{-1}$. 
These quantities are explained in more detail in 
Dickey {\it et al.} (\cite{Dickey_etal_2009}).

Figure 2 shows the ratio of the quantities shown on the two panels of
figure 1, $n_H$ divided by $\frac{n_H}{\left< T_{sp} \right>}$.  Now the units are simply 
Kelvins, as the quantity on the y-axis is simply the mean
spin temperature, $\left< T_{sp} \right>$.  The values are remarkably
constant with radius, $R_G$, in the three surveys, with log($\left< T_{sp} \right>$)
in the narrow range 2.3 to 2.6, i.e 200 to 400 K.  Note that this is not 
necessarily the physical temperature of the gas, since generally there is
a blend of cool clouds (temperature $T_c \sim$50 K typically) and warm
gas ($T_{sp} \gg 500$ K typically) in each spectral channel.

The constancy of the mean spin temperature, $\left< T_{sp} \right>$, with Galactic
radius suggests that the mixture of warm (WNM) and cool (CNM) neutral media is
also fairly constant.
This depends on the cool phase temperature, $T_c$, since the fraction of the
H{\tt I} in the cool phase, $f_c \equiv  \frac{n_{CNM}}{n_{WNM} + n_{CNM}}$, is given by
$f_c = \frac{T_{c}} {\left< T_{sp} \right>}$.  Here the warm phase temperature
is assumed to be much greater than $T_c$, so that its value is unimportant.
This is approximately true in the solar neighborhood, where most of the warm phase
gas is 3000 to 10,000 K (Kulkarni and Heiles \cite{Kulkarni_Heiles_1988}).
The cool phase temperature, $T_c$, is certainly not constant, as there is a 
wide range of temperatures in the H{\tt I} clouds, from below 20K all the way up to the WNM
range (see the article by Troland in this volume), but there is a peak
in the temperature distribution in the range 40 to 60 K (Dickey {\it et al.}
{\cite{Dickey_etal_2003}, Heiles and Troland 
\cite{Heiles_Troland_2003}).  It may be that the peak of this distribution,
which we identify as $T_c$,
changes with $R_G$, and that $f_c$ also changes, so that their ratio, $\left< T_{sp} \right>$,
stays constant.  But a simpler interpretation of the results on figure 2 is that
both $T_c$ and $f_c$ are roughly independent of $R_G$ outside the solar circle.

\section{Conclusions}

The robust nature of the mean spin temperature, $\left< T_{sp} \right>$, is 
an unexpected result.  Somehow conditions even in the far outer disk of the
Milky Way support a mixture of cool phase (CNM or diffuse clouds) and warm
phase (WNM or inter-cloud medium, although much of it is found in and around the 
diffuse clouds) that is not very different from that in the solar neighborhood,
for which $f_c \sim 0.25$ (Kulkarni and Heiles \cite{Kulkarni_Heiles_1987},
Mebold {\it et al.} \cite{Mebold_etal_1997}).  Since the existence of the CNM
in heating-cooling equilibrium with the WNM requires that the pressure be above
the two phase threshold, and the gravitational potential of the stellar disk
at these large values of $R_G$ will not pressurize the medium sufficiently,
the presence of the cool phase is indirect evidence for large scale departures
from pressure equilibrium.  One cause could be converging flow patterns in
the gas, perhaps induced by tidal interactions or by the effect of spiral arms
in the inner disk.  The morphology of the cool 
clouds, which appear in very large complexes, 500 pc or more in diameter,
in the far outer Galaxy (Strasser {\it et al.} \cite{Strasser_etal_2007})
suggests that they result from gas dynamics on very large scales.

{}

\end{document}